\documentclass[aps,prd,preprintnumbers,superscriptaddress,nofootinbib,floatfix,twocolumn,notitlepage]{revtex4-2}
\usepackage{graphicx}  
\usepackage{dcolumn}   
\usepackage{bm}        
\usepackage{epsfig,amsmath,amssymb,verbatim,mathrsfs,array,layout,textcomp,amssymb,latexsym,slashed}
\usepackage{xcolor}
\usepackage[colorlinks=true,citecolor=blue,urlcolor=blue,linktocpage=true,
linkcolor=blue]{hyperref}
\usepackage[utf8]{inputenc}
\usepackage{multirow}
\usepackage{xspace}
\usepackage{booktabs} 
\usepackage{siunitx} 
\usepackage{tabularx} 
\usepackage{physics} 
\usepackage{dsfont} 
\usepackage{physunits} 

\newcommand{\cL}{\mathcal{L}}

\newcommand{\cO}{\mathcal{O}}

\newcommand{\eg}{\textit{e.g.}}
\newcommand{\ie}{\textit{i.e.}}

\newcommand{\alphaSNP}{\alpha_{\phi}}
\newcommand{\alphaVNP}{\alpha_{X}}

\newcommand{\gS}{g_{\rm S}}
\newcommand{\gP}{g_{\rm P}}
\newcommand{\gV}{g_{\rm V}}

\newcommand{\eEDM}{\textit{e}EDM}

\newcommand{\req}{r_{\rm eq}}

\newcommand{\be}{\begin{equation}}
\newcommand{\ee}{\end{equation}}
\newcommand{\bea}{\begin{eqnarray}}
\newcommand{\eea}{\end{eqnarray}}

\newcommand{\molecule}{$\text{CHDBrI}^+\xspace$}

\definecolor{light_blue}{rgb}{0.15, 0.35, 0.9}

\definecolor{Ecolor}{HTML}{008000}
\definecolor{Bcolor}{HTML}{c8ab37}
\definecolor{edm}{HTML}{c83771}

\definecolor{onlyP}{HTML}{003f5c}
\definecolor{onlyN}{HTML}{bc5090}
\definecolor{bothPN}{HTML}{ffa600}

\newcommand{\caseA}{\textcolor{onlyP}{(a)}}
\newcommand{\caseB}{\textcolor{onlyN}{(b)}}
\newcommand{\caseC}{\textcolor{bothPN}{(c)}}
\newcommand{\caseD}{(d)}

\newcommand{\Hrot}{H_{\rm rot}}
\newcommand{\HSN}{H_{\rm SN}}
\newcommand{\Hhfs}{H_{\rm HFS}}

\usepackage[inkscapeformat=png]{svg}

\begin{document}

\title{Constraining P and T Violating Forces with Chiral Molecules}
 
\author{Chaja Baruch}
\email{chajabaruch@campus.technion.ac.il}
\affiliation{Physics Department, Technion -- Israel Institute of Technology, Haifa 3200003, Israel}

\author{P.~Bryan Changala}
\email{bryan.changala@cfa.harvard.edu}
\affiliation{Center for Astrophysics $\vert$ Harvard \& Smithsonian, Cambridge, MA 02138, USA}

\author{Yuval Shagam}
\email{yush@technion.ac.il}
\affiliation{Schulich Faculty of Chemistry, The Helen Diller Quantum Center and the Solid State Institute,
Technion-Israel Institute of Technology, Haifa, 3200003, Israel}

\author{Yotam Soreq}
\email{soreqy@physics.technion.ac.il}
\affiliation{Physics Department, Technion -- Israel Institute of Technology, Haifa 3200003, Israel}

\begin{abstract}
New sources of parity and time reversal violation are predicted by well motivated extensions of the Standard Model and can be effectively probed by precision spectroscopy of atoms and molecules. 
Chiral molecules have distinguished enantiomers which are related by parity transformation. 
Thus, they are promising candidates to search for parity violation at molecular scales, yet to be observed.   
In this work, we show that precision spectroscopy of the hyperfine structure of chiral molecules is sensitive to new physics sources of parity and time reversal violation. 
In particular, such a study can be sensitive to regions unexplored by terrestial experiments of a new chiral spin-1 particle that couples to nucleons.
We explore the potential to hunt for time reversal violation in chiral molecules and show that it can be a complementary measurement to other probes. 
We assess the feasibility of such hyperfine metrology and project the sensitivity in \molecule{}.
\end{abstract}

\maketitle

\section{Introduction}
\label{sec:Intro}

Discrete space-time symmetries such as parity~(P), charge-conjugation~(C) and time-reversal~(T) play an important role in fundamental physics. 
Well-known examples are the discovery of P violation~(PV) in Cobalt-60 $\beta$-decay and of CP violation~(CPV) in kaon decays, both of which were milestones in the building of the Standard Model~(SM).
However, despite its unparalleled success, the SM cannot be considered a complete description of Nature. 
Thus, new degrees of freedom that extend it are required, see \eg~\cite{EuropeanStrategyforParticlePhysicsPreparatoryGroup:2019qin}.
Moreover, new physics sources of PV and CPV are predicted in many beyond SM~(BSM) scenarios and are, for instance, motivated by Baryogenesis.  

Precision atomic and molecular spectroscopy is a powerful tool to probe PV, CPV, and T violation~(TV) effects from SM and new physics sources \eg,~\cite{Pospelov:1997uv,Wood:1997zq,Ginges:2003qt,Chupp:2017rkp,Gaul:2023hdd,Roberts:2014cga,Stadnik:2017hpa,Flambaum:2019ejc,Roussy:2022cmp,Caldwell:2022xwj,ACME:2018yjb} and see~\cite{Safronova:2017xyt} for a review. 
To date, PV is observed at atomic scales and smaller, but it is yet to be observed in molecular systems. 
Chiral molecules, which are polyatomic molecules with two distinguishable enantiomers, left~(L) and right~(R), are promising candidates to observe PV at molecular scales, induced by the electron-nuclei weak interaction~\cite{Letokhov:1975zz,PhysRevLett.83.1554}.
Recent developments are expected to reach the SM PV for the first time in vibrational  spectroscopy \eg,~\cite{quack2015,Cournol_2019,Erez:2022ind,Fiechter2022,leibrandt2024,Lee:2024mrp,Augenbraun:2020wlo} as well as nuclear magnetic resonance spectroscopy \cite{Budker2017}. 
In other attempts using rotational spectroscopy~\cite{satterthawaite2022,Sahu2024}, the magnitude of the SM effect is typically below the current achievable precision.
Chiral molecules may be utilized as a dark matter detectors by a comparison of the L and R vibrational spectra~\cite{Gaul:2020bdq}.
In this work, we investigate the potential of precision rotational and vibrational spectroscopy of chiral molecules to serve as a probe of new physics sources of PV and CPV without relying on the dark matter cosmological abundance. 

New physics sources of PV can result from electron-electron, electron-nucleon or nucleon-nucleon interactions. 
The first two can be probed in electronic and/or atomic systems, \eg~\cite{Bouchiat:2004sp}, while the latter cannot.
Thus, molecules are uniquely sensitive to new nucleon-nucleon interactions with a typical range of an angstrom.
Since a parity transformation relates the two molecular enantiomers, a comparison of them is a direct probe of PV, eliminating the need for external electric or magnetic fields. 
Moreover, the SM weak interaction contribution to the molecular rotational structure is suppressed~\cite{Letokhov:1975zz}.
Therefore, the difference between the L and R hyperfine spectra of chiral molecules is an excellent probe of new PV nucleon-nucleon interactions with negligible SM contribution. 
Thus, there is no need for any sophisticated SM inputs. 
As we show below, precision spectroscopy of chiral molecules can potentially probe parameter space unexplored by terrestrial experiments of a new PV spin-1 particle that couples to neutrons, or to neutrons and protons with opposite coupling. 

CPV or TV (we use them interchangeably as we assume CPT) in molecules originates from several BSM sources, among them electron and nucleon interaction, the electron EDM~($e$EDM), modification of the pion coupling and others, see \eg,~\cite{Pospelov:2005pr,Engel:2013lsa,Prosnyak:2023duq} for more details.  
Here we focus on the nucleon-nucleon interaction and show that in order to hunt for this effect external magnetic and electric fields are required. 
The current bounds from the reinterpretation of the $e$EDM in diatomic molecules~\cite{Baruch:2024frj} are stronger than the expected sensitivity for a new CPV spin-0 state. 
However, the proposed measurement is complementary to the current $e$EDM searches. 

The rest of the paper is organized as follows. 
We introduce our benchmark models in Section~\ref{sec:PCPVforces}.
In Section~\ref{sec:AngMomentum} we analyze the angular momentum hyperfine structure of chiral molecules relevant to new physics searches. 
In Section~\ref{sec:Molecproj}, we give the projection for the \molecule{} molecule, which has been discussed in the context of searches for  SM PV via vibrational spectroscopy \cite{Landau:2023drt,Erez:2022ind,Edu2023}. 
We conclude in Section~\ref{sec:conclusion}.
The Appendix contains technical details which are relevant to this work. 

\section{P- and CP-violating forces}
\label{sec:PCPVforces}

In this section we present our benchmark models for P- and CP-violating forces.
For PV we consider a new spin-1 particle, $X_\mu$, with both vector-and axial-vector couplings to the SM fermions. 
The resulting low energy potential includes a PV part, denoted as $AV$, see \eg~\cite{Dobrescu:2006au,Fadeev:2018rfl}. 
For CPV (or TV), we consider a new spin-0 particle, $\phi$, with both scalar- and pseudo-scalar couplings to the SM fermions, which can induce a CP-violating potential. 
We denote the low energy CPV potential as $PS$, see \eg~\cite{Dobrescu:2006au,Fadeev:2018rfl}.

\subsection{Benchmark parity violating spin-1 model}
\label{sec:PVspin1}

We consider a spin-1 particle $X_\mu$ with both vector and axial-vector couplings to fermions, which induces a parity-violating effect at the molecular scale. 
The effective couplings between $X_\mu$, the nucleons, $N=n,p$, and the electrons are given by 
\begin{align}
    \label{eq:LintVA}
    \cL_{\rm int} 
    \subset   
    \sum_{\psi=e,p,n}\!\!\!
    X^\mu \left(  g_V^X  \overline{\psi} \gamma_\mu \psi  
    +  g_A^X \overline{\psi}\gamma_\mu \gamma_5 \psi  \right)\, .
\end{align}
The effective couplings in Eq.~\eqref{eq:LintVA} are constrained both by terrestrial and astrophysical observations. 
At the molecular scale, the relevant constraints are those for $X_\mu$ with $m_X \lesssim 10\,\eV[k]$.
Regarding terrestrial constraints, the strongest bound on $g_A^N$ is set using the J-coupling interaction in HD $g_A^p(g_A^p + g_A^n) \lesssim 1.4 \times 10^{-18}$ for $m_X\ll \eV[k]$~\cite{Ledbetter:2012xd}. 
The vector interactions $\gV^n$ are constrained by neutron scattering~\cite{Nesvizhevsky:2007by,Kamiya:2015eva,Haddock:2017wav,Heacock:2021btd} and molecular vibrational modes in HD \eg,~\cite{SchillerHD1,SchillerHD2} $g_V^p (g_V^p+g_V^n) \lesssim 2.9\times 10^{-10}$ for $m_X\ll\eV[k]$.
(We do not consider bounds from anomalous currents, \eg~\cite{Dror:2017ehi,Dror:2017nsg}, which are model dependent.)
The P-violating proton-nucleon interaction is also constrained by the nuclear anapole momemt of $^{133}$Cs $g_p^A g_N^V \leq 6.0 \times 10^{-8}$~\cite{Dzuba:2017puc} , see~\cite{PhysRevResearch.5.013191} for a new method to measure the TI anapole moment.
For $m_X \lesssim 10\,\eV[k]$, there are stringent astrophysical bounds on $g_V^N$ obtained from stellar cooling~\cite{Hardy:2016kme}.

The effective P-violating potential between nuclei $i$ and $j$ due to the interaction of Eq.~\eqref{eq:LintVA}  is given by~\cite{Dobrescu:2006au,JacksonKimball:2014vsz,Fadeev:2018rfl}
\begin{align}
    \label{eq:Vva}
    V_{AV}(r) 
=   V^{(1)}_{AV}(r) + V^{(2)}_{AV}(r)  \, , 
\end{align}
with
\begin{align}
    V^{(1)}_{AV}(r)
&=  \alphaVNP^{ij} 
    \frac{\left(\vec{\sigma}_i \times \vec{\sigma}_j \right)\cdot \hat{r}_{ij}}
    {2\overline{m}_N} 
    \left(\frac{1}{r} + m_X \right) \frac{e^{-r m_X}}{r} \, \nonumber \\ 
    V^{(2)}_{AV}(r)
&=  \alphaVNP^{ij}  
    \frac{\vec{\sigma}_i}{2} \cdot \left\{\frac{\vec{p}_i}{m_i}- \frac{\vec{p}_j}{m_j}, \frac{e^{-rm_X}}{r}\right\} \, ,
\end{align}
where $\vec{\sigma}_j$ are the Pauli matrices that follow the spin of the valence nucleon, $\overline{m}_N = 939\,\eV[M] $ is the average nucleon mass and $\vec{r}$ is the inter-nuclear axis of the molecule. 
The new physics interaction strength is defined as
\begin{align}
    \label{eq:alphaNP}
    \alphaVNP^{ij} \equiv 
    -\frac{1}{4\pi} 
    (Z^i g_V^p + N^i g_V^n)
    (B^j_p g_A^p  + B^j_n g_A^n) \,,
\end{align}
where $Z^i\,(N^i)$ is the number of protons\,(neutrons) in nucleus $i$. 
The relation between the nucleus and the nucleon spins, as well as the proton-neutron mass difference, is encoded in $B^j_{n,p}$~\cite{JacksonKimball:2014vsz,Flambaum:2006ip}.
The electron-electron and electron-nuclei potentials can be derived from Eqs.~\eqref{eq:Vva} and~\eqref{eq:alphaNP} by taking $\alpha^{ee}_X=-g^e_V g^e_A/(4\pi)$ and $\alpha^{ej}_X=-g^e_V(B^j_p g_A^p+B^j_n g_A^n)/(4\pi)$ or $\alpha^{je}_X=-(Z^i g_V^p + N^i g_V^n)g_A^e/(4\pi)$ for $ee$ and $eN$, respectively. 
For nucleus-nucleus interactions the $V^{(1)}_{AV}(r)$ part of Eq.~\eqref{eq:Vva} is the dominant contribution, since $V^{(2)}_{AV}(r)$ scales with the nuclei velocity, which is very small, see below.

\subsection{Benchmark spin-0 model}
\label{sec:PSspin0}

We consider a spin-0 particle, $\phi$ with both scalar and pseudoscalar couplings to fermions, which induces a CP-violating potential at the molecular scale.
The effective couplings between $\phi$, the nucleons and the electrons are given by 
\begin{align}
    \label{eq:LintPS}
    \cL_{\rm int} 
    \subset   
    \sum_{\psi=e,p,n}\!\!\!\phi &\left( \gS^\psi  \overline{\psi} \psi 
    + i \gP^\psi \overline{\psi}\gamma_5 \psi  \right)\,.
\end{align}
Such an interaction can be mapped onto UV models, such as \eg, the CPV axion~\cite{Moody:1984ba,DiLuzio:2023lmd,DiLuzio:2023edk} and CPV light scalars that are a result of  relaxion mixing with the SM-Higgs~\cite{Graham:2015cka,Gupta:2015uea,Flacke:2016szy}. 
For a summary of existing constraints, see~\cite{OHare:2020wah}.
As for the P-violating case, the effective couplings of~Eq.~\eqref{eq:LintPS} are constrained both by terrestrial and astrophysical observations. 
We probe interactions at the molecular scale, \ie,  $m_\phi\sim\cO(\eV[k])$. 

Regarding terrestrial observations, the strongest bounds set on $\gP^{n,p}$ come from diatomic \eEDM{} $\gS^{n,p}\gP^{n,p}\leq 1.2\times 10^{-17}$~\cite{Baruch:2024frj}. 
Additional bounds arise from the proton and neutron EDMs, $|d_p|<2.1\times 10^{-25}e\,\cm$~\cite{Sahoo:2016zvr} and $|d_n|<1.8\times 10^{-26}e\,\cm$~\cite{Abel:2020pzs,Pendlebury:2015lrz}, which can be translated following~\cite{DiLuzio:2023edk} to $\gS^p \gP^p < 8.4\times 10^{-10}$ and $\gS^n \gP^n < 1.0\times 10^{-10}$.
As for the spin-1 case, the scalar interactions are constrained by neutron scattering~\cite{Nesvizhevsky:2007by,Kamiya:2015eva,Haddock:2017wav,Heacock:2021btd} and molecular vibrational modes in HD \eg,~\cite{SchillerHD1,SchillerHD2}. For additional bounds, see \eg~\cite{Mostepanenko:2020lqe}.
Additional bounds can be set using the the same J-coupling interaction in HD and H$_2$ molecules as for the $g_A^N$ constraint,  $\gP^p(\gP^n+ \gP^p) \lesssim 6.9 \times 10^{-4}$~\cite{Ledbetter:2012xd}. 

Astrophysical constraints for $m_\phi \lesssim 10\,\keV$ are due to hot neutron stars $(\gP^n,\gP^p) < (1.3,1.5) \times 10^{-9}$~\cite{Buschmann:2021juv}, stellar cooling $\gS^N < 6.5\times 10^{-13}$~\cite{Hardy:2016kme,Bottaro:2023gep}, and SN1987 $\gP^N<6.0\times 10^{-10}$~\cite{Carenza:2019pxu}. 
Additionally, the scalar and pseudoscalar proton couplings contribute to the scalar-photon and pseudoscalar-photon couplings at the one loop level. 
Translating these constraints as in~\cite{Escribano:2020wua} we obtain a bound of $\cO(10^{-16})$ on $g_S^p g_P^p$. 
It is worth noting that both the SN1987 bound and globular cluster bounds~\cite{Dolan:2022kul} have been suggested to be subject to substantial uncertainties, casting doubt on their reliability~\cite{Bar:2019ifz}.
All of these astrophysical bounds can be avoided in models that are subject to environmental effects, see \eg,~\cite{Burrage:2016bwy,Masso:2005ym,Jaeckel:2006xm,DeRocco:2020xdt,Budnik:2020nwz,Bloch:2020uzh} and discussion in~\cite{OHare:2020wah}.

The effective T-violating (which is also PV and CPV) potential between two nuclei $i$ and $j$ is given by~\cite{Dobrescu:2006au,JacksonKimball:2014vsz,Fadeev:2018rfl}
\begin{align}
    \label{eq:Vps}
    V_{PS}(r)
&=  \alphaSNP^{ij} \frac{\vec{\sigma}_j\cdot\hat{r}_{ij}}{2\overline{m}_N} 
    \left(\frac{1}{r} + m_\phi \right) \frac{e^{-r m_\phi}}{r}\,.
\end{align}
The new physics interaction strength is defined as in Eq.~\eqref{eq:alphaNP} but we replace $X\to\phi$, $V\to S$ and $A\to P$.

\section{Angular momentum structure in polyatomic molecules}
\label{sec:AngMomentum}

Chiral molecules are polyatomic molecules that possess a handedness, \ie, their mirror images are distinguishable. 
This asymmetry probes parity-violating energy differences caused by both SM interactions (\eg, $Z$ boson exchange between electrons and nuclei~\cite{Quack2008}) and new physics interactions. 
In this section, we identify hyperfine states of rotating chiral molecules $\vert \Psi_\mathrm{mol} \rangle$ that satisfy two criteria: 
(i)~the expectation values of the new physics interaction potentials, Eq.~\eqref{eq:Vva} and/or \eqref{eq:Vps}, are non-zero, and 
(ii)~the sign of this expectation value can be experimentally switched to enable a differential measurement of the new physics term. 
Meeting these criteria requires an examination of the alignment and orientation of various (rotational and spin) angular momenta and bond vectors within the rotating molecule-fixed frame.

The key factor of the effective $AV$ potential, for example, is the term $\langle \Psi_\mathrm{mol} \vert\left(\vec{\sigma}_i \times \vec{\sigma}_j \right)\cdot \hat{r}_{ij} \vert \Psi_\mathrm{mol} \rangle \equiv \langle \left(\vec{\sigma}_i \times \vec{\sigma}_j \right)\cdot \hat{r}_{ij}  \rangle$. 
As we discuss below, this operator has a non-zero expectation value if the nuclear spins have a fixed relative orientation to each other, not necessarily in the molecule-fixed frame, and can be switched by swapping L and R enantiomers. 
The corresponding factor of the effective $PS$ potential, $\langle \vec{\sigma}_j\cdot\hat{r}_{ij} \rangle$, requires the nuclear spins to be \textit{oriented}, not just aligned, in the molecule-fixed frame using external magnetic and electric fields, but is otherwise also switchable with chiral enantiomers.

\subsection{The effective hyperfine Hamiltonian}
\label{sec:H}

The hyperfine eigenstates are characterized by a minimum of six quantum numbers: 
$N$ the rotational angular momentum, 
$K$ the projection of $N$ on the quantization axis of the molecule, 
$S$ the electronic spin, 
$J=N+S$ the total electronic angular momentum, $I_n$ the nuclear spin of atom $n$, the intermediate sums of electronic and nuclear angular momentum $F_{n-1}=J+I_1+..+I_{n-1}$ and finally $F=N+S+\sum_i I_i$, the total angular momentum. 
Additionally, we define the labels $K_a, K_c$ for the different projection states of an asymmetric top molecule~\cite{Demtrder2005MolecularPT,Herzberg1945MolecularSA}.
The effective hyperfine Hamiltonian is
\begin{align}
    H 
=   \Hrot + \HSN + \Hhfs\,.
\end{align}
The rotational angular momentum is described by $\Hrot$, where we used a rigid rotor approximation
\begin{align}
    \label{eq:Hrot}
    \Hrot 
=   \frac{1}{2} \vec{N} \cdot \mu \cdot \vec{N}\,,
\end{align}
where $\mu$ is the inverse moment-of-inertia tensor. 
We choose the molecule-fixed frame to be the principal axis system $\{x,y,z\}$, in which $\mu$ is diagonal.

The electronic spin of the molecule is coupled to the rotational angular momentum  via
\begin{align}
    \label{eq:HSN}
    \HSN 
=   \vec{S} \cdot \epsilon \cdot \vec{N} \, ,
\end{align}
where $\epsilon$ is the electronic spin-rotation coupling tensor~\cite{VanVleck1951:angmom}.
We consider two hyperfine interactions, a nuclear electric quadrupole term and an electron-nucleus spin-spin interaction $\Hhfs = H_{\rm Q} + H_{\rm SI}$, 
\begin{align}
    H_{\rm Q} &= \sum_i \vec{I}_i \cdot \chi_i \cdot \vec{I}_i \\
    H_{\rm SI} &= \sum_i \vec{S} \cdot A_i \cdot \vec{I}_i\,,
\end{align}
where $\chi_i$ are the nuclear electric quadrupole coupling tensors and $A_i$ are the spin-spin coupling tensors for the electronic and nuclear spins.
In this case $A_i$ is composed of the Fermi contact interaction $a_F^i$ and $T_i$ which is the nuclear spin-dipole coupling.  

\subsection{Parity violating effect}
\label{sec:Pswitch}

For the $AV$ potential of Eq.~\eqref{eq:Vva} to induce a PV contribution to the molecular energy levels, we require a nonzero $\expval{(\vec{\sigma}_i \times \vec{\sigma}_j)\cdot \hat{r}_{ij}}$. 
To obtain this, the nuclear spins must only be mutually oriented in the molecule-fixed frame.
In other words, $\Hrot$ aligns the direction of $\vec{N}$ in the molecule frame, $\HSN$ transfers this alignment to $\vec{S}$, and $H_\mathrm{SI}$ finally transfers it to each $\vec{I}_i$. 
The nuclear electric quadrupoles $\chi_i$ slightly modify the effect. 
Although the individual $\vec{I}_n$ are not \textit{oriented}, their pairwise cross-products are oriented, because they are coupled to a common electronic spin $\vec{S}$.
This means that PV nucleon-nucleon interaction can be probed in chiral molecules \textit{without any external field}, which is similar to the case of SM PV via $Z$-exchange~\cite{Quack2008}.
We consider eigenstates of $H$ (for which $F$ is a good quantum number), and to have a nonzero contribution of the $AV$-potential we require these states to be superposition states of different $\sum_i I_i$ (or equivalently, different $J$).

For a given transition, we define the P switch as the difference between the left and right enantiomer transition energies 
\begin{align}
    \label{eq:DeltaE}
    \Delta E_{\rm PV} 
    \equiv 
    \frac{E_R - E_L}{2}
    \approx
    E^{\rm SM}_{\rm PV} + \sum_{ij}E^{ij}_{\rm PV}\, ,
\end{align}
where $E^{\rm SM}_{\rm PV}$ is the SM PV contribution, which can be neglected in the transitions we are interested in, see below.
The second term is the new physics effect and the sum runs over all nuclei pairs. 
There are three general new physics sources of PV
\begin{align}
    \label{eq:omegaij}
    E_{\rm PV}^{ij}
=   W^{P}_{ee} \alphaVNP^{ee}   
    + W^{P,ie}_{eN} \alphaVNP^{ie} 
    + W^{P,ij}_{NN} \alphaVNP^{ij}  \, ,
\end{align}
\ie, due to electron-electron, electron-nucleon and nucleon-nucleon interactions. Here we focus on $W^{P,ij}_{NN}$ and leave discussion of $W^{P,ie}_{eN}$ and $W^P_{ee}$ to future work. 

We obtain $W^{P,ij}_{NN}$ by first-order perturbation theory and the potential of Eq.~\eqref{eq:Vva}, 
\begin{align}
    \label{eq:WNN_P}
    W^{P,ij}_{NN}
=   \frac{\left\langle V_{AV} \right\rangle}{\alphaVNP^{ij}}
    \approx
    \frac{\expval{\vec{\sigma}_i \times \vec{\sigma}_j}\cdot \hat{r}_{\rm eq}^{ij}}{2\overline{m}_N} \left(\frac{1}{\req^{ij}} + m_\phi \right) \frac{e^{-\req^{ij} m_\phi}}{\req^{ij}} \, ,
\end{align}
where $\req^{ij}$ is the equilibrium distance between the two nuclei. 
In polyatomic molecules, many possible atom pairs contribute to $W^{P,ij}_{NN}$, but their expectation values $\expval{\vec{\sigma}_i \times \vec{\sigma}_j}$ vary drastically between the different molecular eigenstates.

A full analysis of the contributions of the $eN$ and $ee$ interactions is beyond the scope of this work. 
However, for $m_X\gg \eV[M]$ the $VA$ potential is a rescaling of the SM weak interaction. 
Thus, we estimate the sensitivity to the $eN$ contributions as a function of the uncertainty in the SM measurement. 
Assuming that the PV shift in the vibrational modes is consistent with the SM prediction, we project an upper bound of
\begin{align}
    \label{eq:gVngAeProj}
    \frac{\abs{g_V^p g_A^e}}{\sqrt{2} m_X^2} &\leq \Delta  G_{\rm F} \left(\frac{1}{4}-s_W^2\right)\,, \nonumber \\ 
    \frac{\abs{g_V^n g_A^e}}{\sqrt{2} m_X^2} &\leq \Delta G_{\rm F} \left(\frac{1}{2}\right)\,,
\end{align}
where $s_W$ is the sine of the Weinberg angle, $G_{\rm F}\simeq 1.17\times 10^{-5}\,\eV[G]^{-2}$ is the Fermi constant and $\Delta$ is the relative uncertainty in the measurement.

\subsection{Time reversal violating effect}
\label{sec:Tswitch}

For the $PS$ potential of Eq.~\eqref{eq:Vps} to induce a T-violating contribution to the molecular energy levels, we need a nonzero expectation value $\expval{\vec{\sigma}\cdot\hat{r}}$. 
With zero external field, the nuclear spins (or any other angular momentum) are aligned with respect to the molecule frame, but not oriented, so that $\expval{\vec{\sigma}\cdot\hat{r}} = 0$.
We obtain nonzero $\expval{\vec{\sigma}\cdot\hat{r}}$ by applying both an external electric and external magnetic field. 
The $E$-field couples the molecular frame to the lab frame via the molecular electric dipole, and the $B$-field lifts the remaining degeneracy between the orientations. 
Generally, to then see any T-violating effects, the direction of both EM fields must be reversed. 
This is not the case for chiral molecules, where we can reverse the sign of just one of the EM fields, and the second switch is obtained by comparing the transitions of the left- and right-handed enantiomer. 
Thus, we define the double difference 
\begin{align}
    \Delta E_{\rm TV} 
    \equiv 
    \frac{\left(E_L^\uparrow-E_L^\downarrow\right) - \left(E_R^\uparrow-E_R^\downarrow\right)}{4} 
    \approx
    \sum_{ij} E_{\rm TV}^{ij} \, ,
\end{align}
where the arrow indicates the reversal of the direction of the external magnetic field in the lab frame.
The SM TV contribution is negligibly small.

We consider four sources of T-violation in molecules (for other potential sources see \eg~\cite{Pospelov:1997uv,Ginges:2003qt,Chupp:2017rkp,Gaul:2023hdd,Stadnik:2017hpa,Flambaum:2019ejc}), which can be written as
\begin{align}
    \label{eq:omegaij}
    E_{\rm TV}^{ij}
=   W^{T,ij}_{d_e} d_e 
    + W^{T,ie}_{eN} \alphaSNP^{ie} 
    + W^{T}_{ee} \alphaSNP^{ee}
    + W^{T,ij}_{NN} \alphaSNP^{ij}  \, . 
\end{align}
As for the $VA$ case, we focus on $W^{T,ij}_{NN}$ and leave the electron-electron and electron-nucleon interactions for future work. 
Moreover, these interactions can be probed in atomic systems \eg~\cite{Stadnik:2017hpa}.
To estimate the contribution of a possible \eEDM, we need to compute the effective electric field exerted on an electron in the molecule, which is beyond the scope of this work. 
The nucleon-nucleon contribution can be estimated by first-order perturbation theory
\begin{align}
    W^{T,ij}_{NN}
=   \frac{\left\langle V_{\rm SP} \right\rangle}{\alphaSNP^{ij}}
    \approx
    \frac{\langle\vec{\sigma}_j \cdot \hat{r}_{\rm eq}^{ij} \rangle}{2\overline{m}_N} 
    \left(\frac{1}{\req^{ij}} + m_\phi \right) 
    \frac{e^{-\req^{ij} m_\phi}}{\req^{ij}} \, .
\end{align}

We briefly note that, in principle, the TV necessary for non-zero $\expval{\vec{\sigma}\cdot\hat{r}}$ could be obtained instead by preparing non-stationary rotational quantum states that have a non-zero signed projection ($K$) of the rotational angular momentum $N$ in the molecule-fixed frame. 
In such ``gyroscopic'' states, the molecule can rotate with a well-defined orientation in both the molecule-fixed and lab frames. 
The spin-rotation and hyperfine interactions would ultimately transfer this orientation to individual nuclear spins. 
For potential \eEDM measurement in molecules without external fields see~\cite{Zhang:2023sme}.

\section{Projection for \molecule}
\label{sec:Molecproj}

\subsection{Parity violating projection }
\label{sec:PVproj}

%
\begin{table}
    \begin{tabular}{lrr}
    \hline 
    Parameter & CH$_2$$^{79}$BrI$^+$  & CHD$^{79}$BrI$^+$ \\ 
    \hline
    $A$/MHz &  16888.0  &  14781.2   \\
    $B$/MHz &  1109.2   &  1106.7    \\
    $C$/MHz &  1048.3   &  1040.0    \\
    \\[-3mm]
    $\mu_a$/au &  $-$0.170  & $-$0.168   \\
    $\mu_b$/au &  0.825  &  0.824   \\
    $\mu_c$/au &  0   &  $-$0.048   \\
    \\[-3mm]
\hline
    \end{tabular}
    \caption{The rotational constants and permanent dipole moments of CH$_2$BrI$^+$ and CHDBrI$^+$.}
    \label{tab:ch2bri_rot}
\end{table}
\begin{table*}
    \begin{tabular}{l@{\qquad}rrr@{\qquad}rrr}
    \hline 
    Tensor\footnotemark[1] & \multicolumn{3}{c}{CH$_2$$^{79}$BrI$^+$}  & \multicolumn{3}{c}{CHD$^{79}$BrI$^+$} \\ 
    \hline
                      & $-$20323.7   & $-$1515.1  &   0.0           &  $-$17872.6 & $-$1509.9  &    84.4\\ 
     $\epsilon$\footnotemark[2]  & $-$17887.0     & $-$1740.7 &   0.0          &  $-$15640.1 & $-$1724.1  &   100.3\\ 
                      &      0.0         &     0.0        &  74.4 &     935.5 &   107.1  &    67.8 \\ 
    \\[-3mm]
                      & 419.1 &   321.0  &    0.0          &    420.6 &   319.1 &  $-$19.4 \\
     $T$(Br)          & 321.0  &  $-$135.3 &    0.0          &    319.1 &  $-$137.4 &   $-$8.9 \\
                      &   0.0         &     0.0         & $-$283.8  &    $-$19.4 &    $-$8.9 & $-$283.2 \\
      \\[-3mm]     
                       &      293.3 &     $-$302.8 &        0.0  &      291.8 &     $-$302.6 &       16.8 \\
     $\chi$(Br)        &     $-$302.8 &      177.6 &        0.0  &     $-$302.6 &      176.9 &      $-$37.5 \\
                       &        0.0 &        0.0 &     $-$470.9  &       16.8 &      $-$37.5 &     $-$468.7 \\

    \\[-3mm]

            &      565.2 &     $-$835.8 &        0.0  &      561.2 &     $-$835.7 &       47.6 \\
  $T$(I)    &     $-$835.8 &       31.1 &        0.0  &     $-$835.7 &       33.0 &      $-$35.9 \\
            &        0.0 &        0.0 &     $-$596.2  &       47.6 &      $-$35.9 &     $-$594.2 \\
    \\[-3mm]
             &    $-$1381.1 &     $-$190.3 &        0.0  &    $-$1382.0 &     $-$187.6 &       14.6 \\
 $\chi$(I)   &     $-$190.3 &     $-$478.7 &        0.0  &     $-$187.6 &     $-$469.8 &      136.4 \\
             &        0.0 &        0.0 &     1859.8  &       14.6 &      136.4 &     1851.8 \\

\\[-3mm]

                               &      872.4 &      925.8 &     $-$672.6  &      875.2 &      885.7 &     $-$734.2 \\
 $T$(H\footnotemark[3]) $\times 10^3$      &      925.8 &     2220.7 &    $-$1805.7  &      885.7 &     1987.8 &    $-$2102.1 \\
                               &     $-$672.6 &    $-$1805.7 &    $-$3093.0  &     $-$734.2 &    $-$2102.1 &    $-$2863.0 \\

                         &&&  &      134.8 &      148.7 &       94.7\\
$T$(D) $\times 10^3$ &&&  &      148.7 &      369.7 &      227.4\\
                         &&&  &       94.7 &      227.4 &     $-$504.5\\
                            &&& &      $-$97.6 &        4.0 &        6.5\\
$\chi$(D) $\times 10^3$ &&& &        4.0 &       $-$4.3 &      142.0\\
                            &&& &        6.5 &      142.0 &      101.9\\
  \hline 
    \end{tabular}
    \caption{The spin-rotation and anisotropic hyperfine coupling constants of CH$_2$BrI$^+$ and CHDBrI$^+$ (in MHz).\label{tab:ch2bri_aniso}}
    \footnotetext[1]{For each tensor quantity, the rows and column are listed in axis order $abc$.}
    \footnotetext[2]{The spin-rotation tensor indices are ordered here as $\epsilon_{ij}(S_i N_j + N_j S_i)/2$. The values for CHD$^{79}$BrI$^+$ are derived from those of the parent isotopologue assuming pure spin-orbit contributions~\cite{Brown1980:spinrot_iso}.}
    \footnotetext[3]{The H values are listed only for the H nucleus that remains unsubstituted in the deuterated isotopologue.}
\end{table*}
\begin{table}   
    \begin{tabular}{cc} 
    \hline 
    Nucleus & $a_F$ (MHz)\\ 
    \hline 
    $^{79}$Br & 167.0 \\ 
    I & 256.7 \\ 
    H & $-$1.998 \\ 
    D & $-$0.307 \\ 
    \hline 
    \end{tabular}
    \caption{The isotropic hyperfine coupling constants of CH$_2$BrI$^+$ and CHDBrI$^+$.}
    \label{tab:ch2bri_isohfcc}
\end{table}
\begin{table*}[]
    \centering
    \begin{tabular}{c|c|c|c|c|c|c}
    & & & \multicolumn{2}{c|}{state 1}  & \multicolumn{2}{c}{state 2}  \\ 
       & $(i,j)$  & $E_{1\rightarrow2}$ [\Hz[G]] & $\bra{N, K_a, K_c, J, F_1, F_2, F_3, F}$  & $\expval{\sigma_i\times\sigma_j}\cdot \hat{r}_{\rm eq}^{ij}$  & $\ket{N, K_a, K_c, J, F_1, F_2, F_3, F}$ & $\expval{\sigma_i\times\sigma_j}\cdot \hat{r}_{\rm eq}^{ij}$ \\
       \hline 
       \hline 
       \caseA,\caseB,\caseC & (D, I)   & 3.34 & $\bra{1,1,0,\frac{3}{2},1,1,\frac{5}{2},0}$ & 0.49 & $\ket{1,0,1,\frac{1}{2},1,2,\frac{7}{2},1}$ & 0.12 \\
       \caseD & (H, I)  & 3.99 & $\bra{2,2,0,\frac{5}{2},3,3,\frac{5}{2},0}$ & 0.23  & $\ket{1,0,1,\frac{3}{2},1,2,\frac{3}{2},1}$ & 0 \\
    \end{tabular}
    \caption{Summary of the optimal transitions to use for probing the PV effect of Eq.~\eqref{eq:Vva} on the rovibrational energy levels of \molecule. 
    We denote the atom pair that gives the leading contribution to the new physics effect $(i,j)$ for each of the benchmarks outlined above, the optimal transition $E_{1\to 2}$, the quantum numbers (with the nuclear spins organized as (H, D, Br, I)) and the expectation value of the polarization for both the initial and final state. 
    }
    \label{tab:P_transitions}
\end{table*}
\begin{table}
    \label{tab:ch2bri_geo}
    \begin{tabular}{lrrr}
    \hline 
        Atom  &     $a$ & $b$ & $c$ \\ 
        \hline
        C &   0.51 &  1.27 & 0       \\
        I &  $-$1.20 & $-$0.07 & 0         \\
        $^{79}$Br &   1.84 & $-$0.14 & 0 \\
        H &   0.54 &  1.85 & $-$0.92 \\
        H &   0.54 &  1.85 &  0.92 \\
        \hline 
    \end{tabular}
    \caption{The Cartesian coordinates of CH$_2$BrI$^+$ and \molecule{} in their PAS, in angstrom.}
\end{table}
\begin{figure*}\label{fig:P_switch_bound}
    \centering 
    \includegraphics[width=0.48\textwidth]{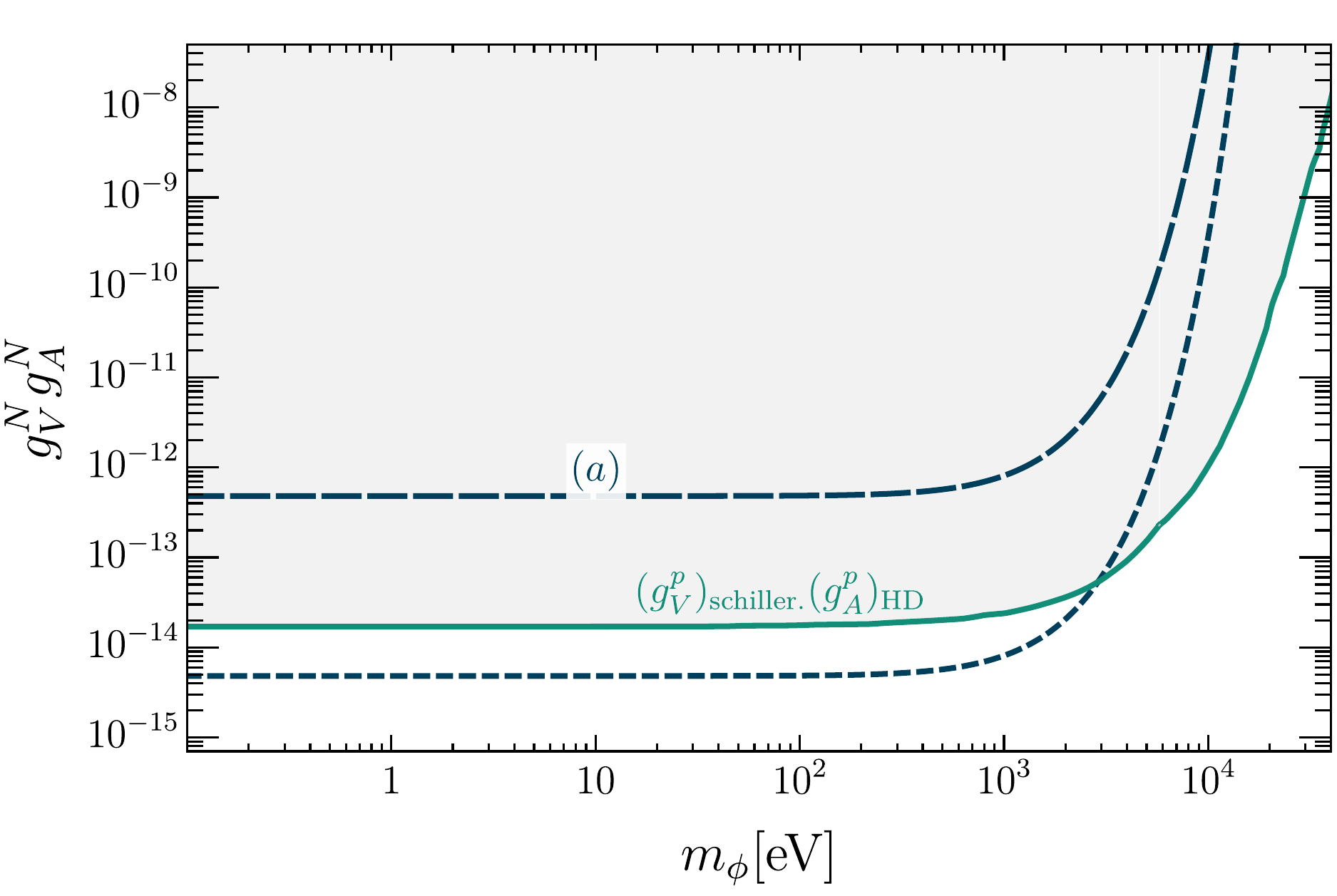}
    \includegraphics[width=0.48\textwidth]{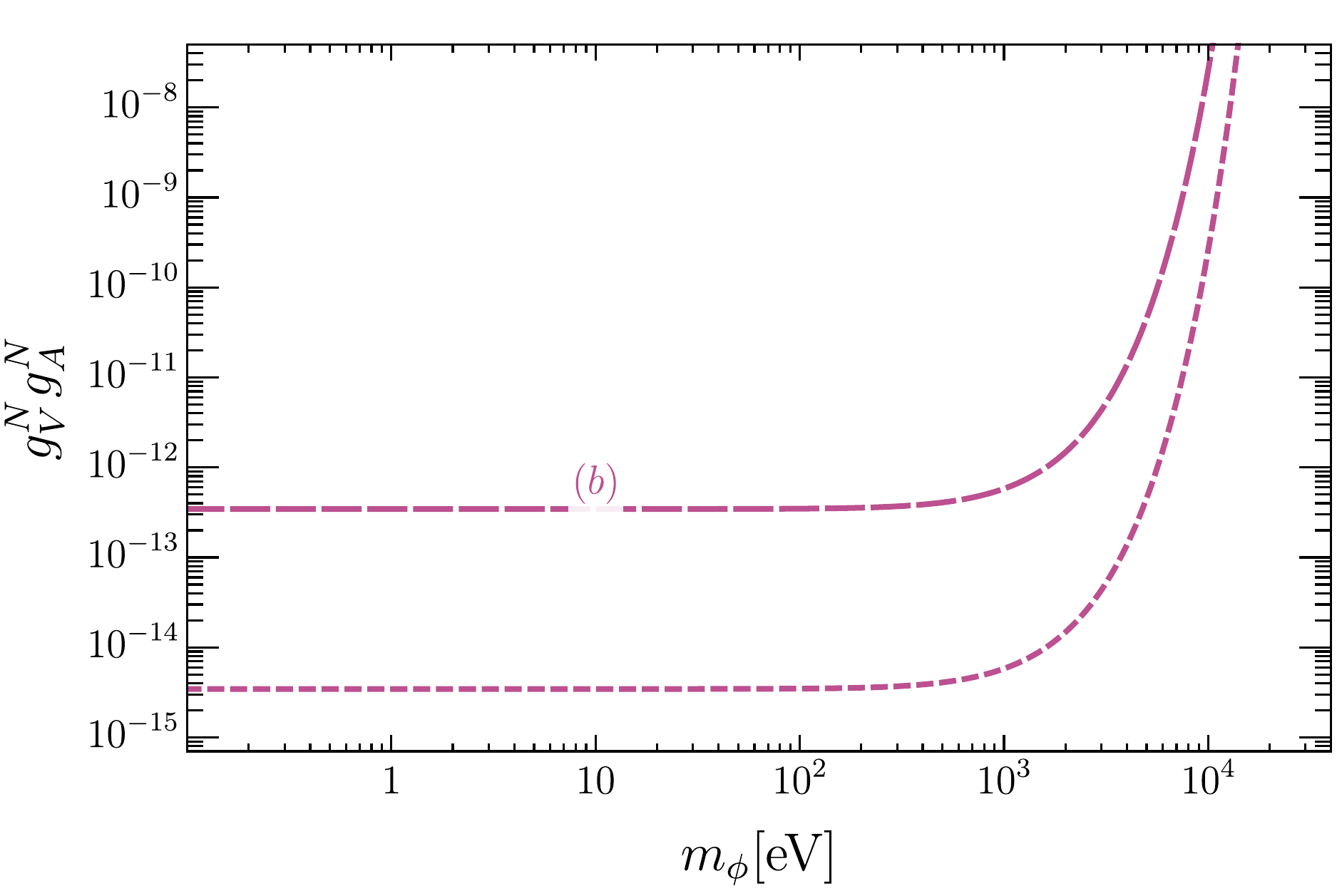}
    \includegraphics[width=0.48\textwidth]{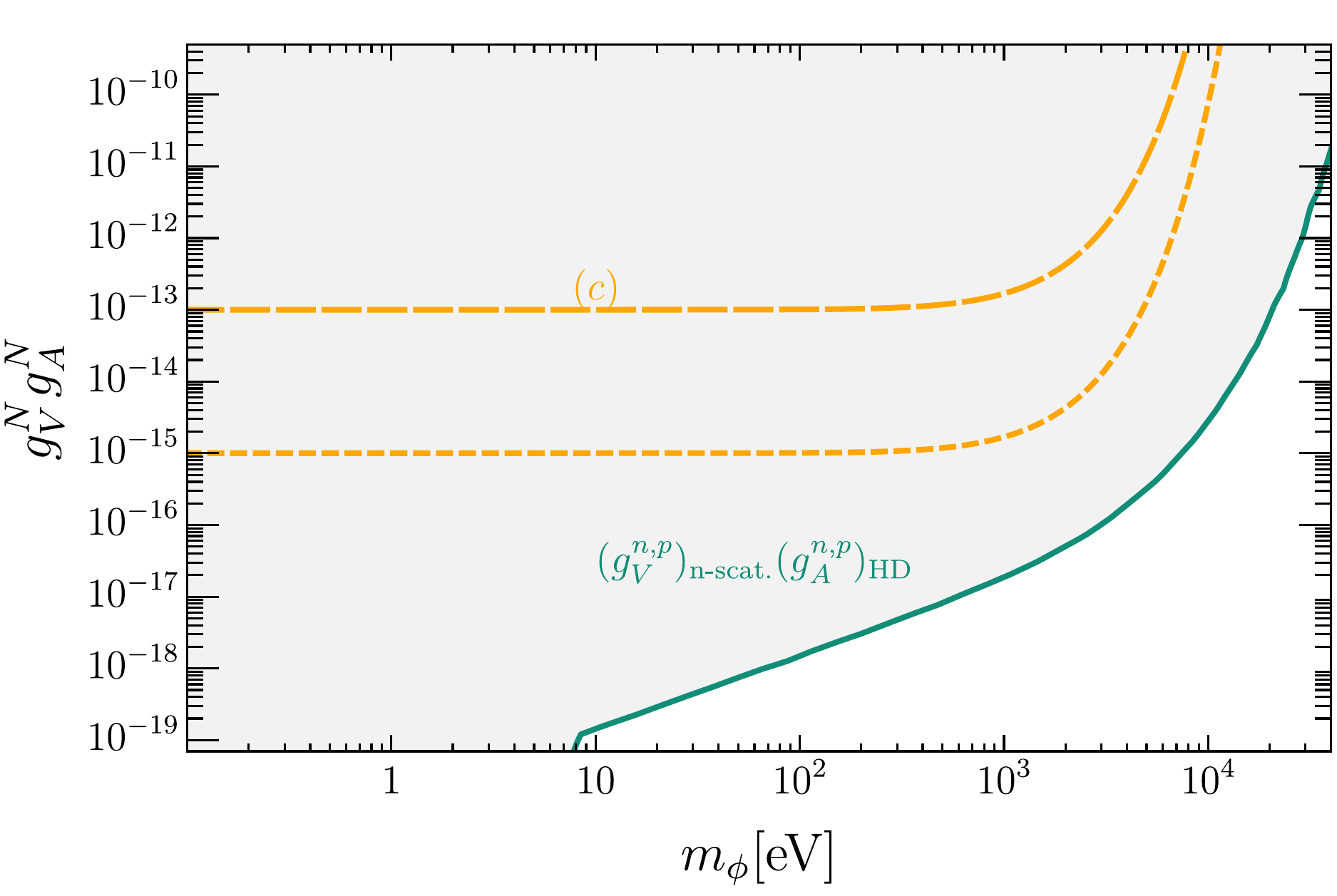}
    \includegraphics[width=0.48\textwidth]{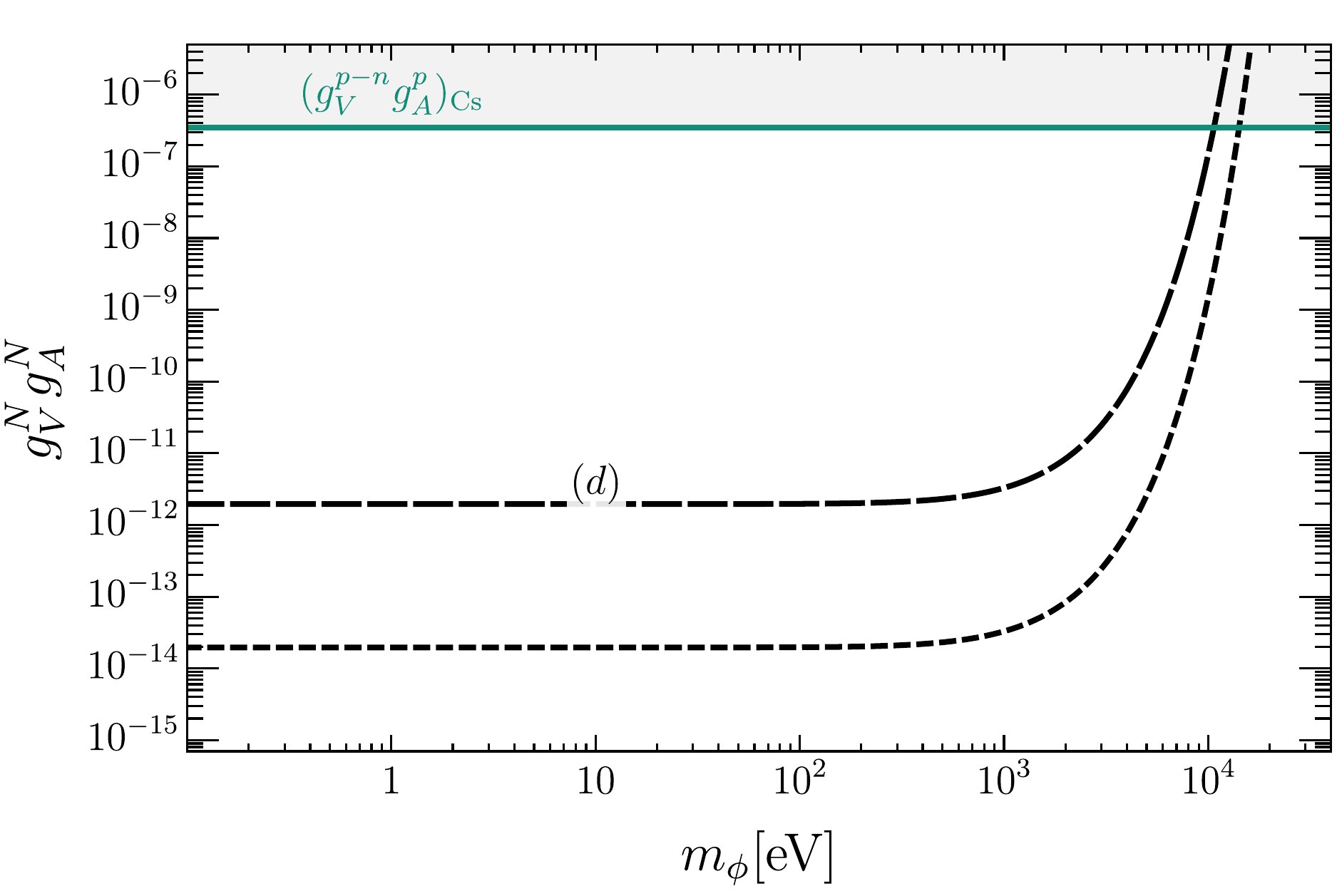}
    \caption{Constraint on the V-A couplings for the different benchmark scenarios, when assuming an accuracy of the measurement of $0.1\Hz$ (long dashes) and $1\Hz[m]$ (short dashes). For \caseA ($g_V^n=g_A^n=0$), the existing constraint is obtained by combining the bounds on $g_V^p$ from HD vibrational modes~\cite{SchillerHD1,SchillerHD2} and the $g_A^p$ bound from the J-coupling in HD~\cite{Ledbetter:2012xd}. For \caseB ($g_V^p=g_A^p=0$) there are no terrestrial experiments that constraint the coupling. For \caseC ($g_V^p=g_V^n, g_A^p=g_A^n$), the existing constraint in the combined bound of neutron scattering~\cite{Nesvizhevsky:2007by,Kamiya:2015eva,Haddock:2017wav} and J-coupling in HD~\cite{Ledbetter:2012xd}. The strongest terrestrial constraint for \caseD ($g_V^p=-g_V^n, g_A^p=-g_A^n$) is due to the nuclear anapole moment of $^{133}$Cs~\cite{Dzuba:2017puc}. }
\end{figure*}

In this section we estimate the sensitivity of precision hyperfine spectroscopy of \molecule{} to probe PV from new physics. 
The proposal for producing the \molecule{} in an internally cold state via photoinization \cite{Landau:2023drt} makes it suitable for precision searches in rotational spectroscopy in addition to the vibrational spectroscopy applications \cite{Edu2023,Erez:2022ind,Landau:2023drt}. 

The rotational constants for \molecule{} were calculated as part of this work and are given in Tab.~\ref{tab:ch2bri_rot} and are in agreement with those computed in Ref.~\cite{Landau:2023drt}.
The calculated hyperfine coupling matrices, including the nuclear quadrupole-electric field gradient coupling tensor $\chi_i$, the electron spin-rotation coupling tensor $\epsilon$ , and the electronic-nuclear spin-dipole coupling tensor $T_i$, are given in Tab.~\ref{tab:ch2bri_aniso}.  
The Fermi contact parameters ($a_F$) are presented in Tab.~\ref{tab:ch2bri_isohfcc}. 
The structural and electronic properties were calculated using the CFOUR quantum chemistry code \cite{Matthews2020:cfour,CFOUR}. 
Details about the methods used are given in Appendix~\ref{app:Qcomp}. 
Values in the tables are given for both CH$_2$BrI$^+$ and the chiral isotopologue {\molecule}, which is the molecule of interest.

For simplicity, we assume that there are only nucleon-nucleon new physics interactions. 
Their dominant effect is given in Eq.~\eqref{eq:WNN_P}, where we consider only $V_{AV}^{(1)}(r)$ and neglect $V_{AV}^{(2)}(r)$. 
The effect of $V_{AV}^{(2)}(r)$ is sub-leading since it is proportional to the momentum of the nuclei.  
In \molecule, the most energetic vibrational mode is the C-H stretch mode, which has a frequency of $91\,\Hz[T]$~\cite{Landau:2023drt} and leads to a proportionality factor $p/\mu \sim \cO(10^{-5})$ compared to $V^{(1)}_{AV}(r)$, where $p$ is the momentum in the rest frame and $\mu$ the reduced mass of the C-H system.

So far the precision in spectroscopy of \molecule{} has only been estimated for vibrational transitions at below $0.1\,\Hz$ for C-H stretch and even better for the C-H wag mode due to the seconds long natural lifetime~\cite{Erez:2022ind,Landau:2023drt,Edu2023}. 
In the rotational spectroscopy needed for the BSM effects discussed in this work, the natural lifetime of rotational transitions will be much longer. 
Thus, the coherence time is expected to be limited by magnetic field noise due to the non-zero electronic spin of the molecule and black body radiation. 
In the proof of concept experiment in Satterthwaite et al.~\cite{satterthawaite2022} a statistically limited $\sim0.7\,\Hz$ precision was achieved between R and L 1,2-propanediol neutral chiral molecules, while in HfF$^+$ molecular ions a precision of $\sim 20\Hz[\mu]$ was reached~\cite{Roussy:2022cmp}. 
It is challenging to estimate the precision on \molecule{} due to the early stages of the experiment.
Therefore, in this work we consider two cases: a conservative case of $0.1\,\Hz$ and an aggressive case of $1\,\Hz[m]$. 
As we progress in production and trapping of \molecule{} molecules, these estimates will become more refined.

The SM PV shift is typically several orders of magnitude smaller for rotational transitions than in the vibrational transitions \eg~\cite{satterthawaite2022}. 
However, an interesting aspect of \molecule{} is that is only isotopically chiral, such that the two enantiomers have a nearly identical electronic configuration. 
Despite this, the molecule exhibits surprisingly large PV shifts in some of its vibrational transitions~\cite{Edu2023}. 
For rotational transitions the SM PV likely scales in similar fashion to other molecules, but we leave detailed calculations for future work.

In our analysis we consider four benchmark scenarios for $X$-nucleon interactions:
\caseA~only protons $g_V^n=g_A^n=0$;
\caseB~only neutrons $g_V^p=g_A^p=0$;
\caseC~protons and neutrons equally $g_V^p=g_V^n, g_A^p=g_A^n$; 
and \caseD~protons and neutrons oppositely $g_V^p=-g_V^n, g_A^p=-g_A^n$.
For each scenario, we have computed which rovibrational transition is most sensitive to the effect of Eq.~\eqref{eq:Vva}. 
The relevant transitions and expectation values are detailed in Tab.~\ref{tab:P_transitions}. 
The Z-matrix elements of \molecule{} can be found in Tab.~\ref{tab:ch2bri_geo}. 
We choose the inital state of the transition used such that the new physics effect is maximized, and determine the transition with the largest intensity from that state using PGOPHER~\cite{WESTERN2017221}.
The quantum numbers defining the state are those of the main contribution to the relevant eigenstate, but in reality, the eigenstates are superpositions of different quantum numbers.
We obtain the expectation values $\expval{\sigma_i\times\sigma_j}$ by diagonalizing the Hamiltonian for the case where only the nuclei $(i,j)$ have spin. 
We verify that the correction from the inclusion all four nuclear spins is small by checking the change in the energy state in PGOPHER.
With this, we obtain
\begin{align}
    g_V^p g_A^p &\leq 4.8 \times 10^{-13} \nonumber \\
    g_V^n g_A^n &\leq 3.4 \times 10^{-13} \nonumber \\ 
    g_V^{p,n} g_A^{p,n} &\leq 1.0 \times 10^{-13} \nonumber \\ 
    g_V^{p,-n} g_A^{p,-n} &\leq 2.0 \times 10^{-12} \,,
\end{align}
for $m_X \lesssim \eV[k]$ and an accuracy of $0.1 \Hz$. 
The mass-dependent constraints are shown in Fig.~\ref{fig:P_switch_bound}.
For \caseB{}, to the best of our knowledge there exist no other terrestrial constraint on the couplings. This is because any probes of these couplings done using HD molecules are not valid since they probe $g^p\times(g^p+g^n)$. 
For \caseA, in order to be competitive with the existing bound an accuracy of $\sim0.01\,\Hz$ is required. 
For \caseC, the existing terrestrial constraints outperform the ones that can be set using chiral molecules with the current precision. 
For \caseD, in order to be competitive with the existing bound an accuracy of $\cO(\Hz[k])$ is required. 

Regarding the electron-nucleon interaction, we assume that the vibrational mode measurement is compatible with the SM expectation to the $\Delta\sim0.1$ level, $\sim0.1\,\Hz$. 
By using Eq.~\eqref{eq:gVngAeProj} we project the following limit of
\begin{align}
   g_V^p g_A^e &\lesssim 5 \times 10^{-12}  \left(\frac{\Delta}{0.1} \right) \left(\frac{m_X}{10\eV[M]} \right)^2\,, \nonumber \\
   g_V^n g_A^e &\lesssim 8 \times 10^{-11}  \left(\frac{\Delta}{0.1} \right) \left(\frac{m_X}{10\eV[M]} \right)^2\,.
\end{align}
This is comparable with the current bounds from APV~\cite{Bouchiat:2004sp,Dzuba:2017puc}. 
As a proof of concept, we adopt the result of~\cite{PhysRevLett.83.1554} for {CHFClBr} with is consistent with no PV and put an upper limit of $13\,\Hz$ on the L and R difference. 
There is a large uncertainty on the theoretical prediction of the SM PV, Ref.~\cite{Quack2000} predicts $0.06\,\Hz$, while Ref.~\cite{Rauhut:2021soh} predicts $1\,\Hz[m]$.
Conservatively, we use the later SM prediction, thus, $\Delta\sim10^4$ and the resulting upper bounds are
\begin{align}
    [g^p_{V}g^e_A]_{\rm CHFClBr}\lesssim 5\times 10^{-7}\left(\frac{m_X}{10\eV[M]} \right)^2 \, ,
    \nonumber\\
    [g^n_{V}g^e_A]_{\rm CHFClBr}\lesssim 8\times 10^{-6}\left(\frac{m_X}{10\eV[M]} \right)^2 \, ,
\end{align}
which, to the best of our knowledge, are the first bounds on BSM from chiral molecules.

\subsection{Time Violating projection }
\label{sec:TVproj}

%
\begin{figure}
    \centering
    \includegraphics[width=\columnwidth]{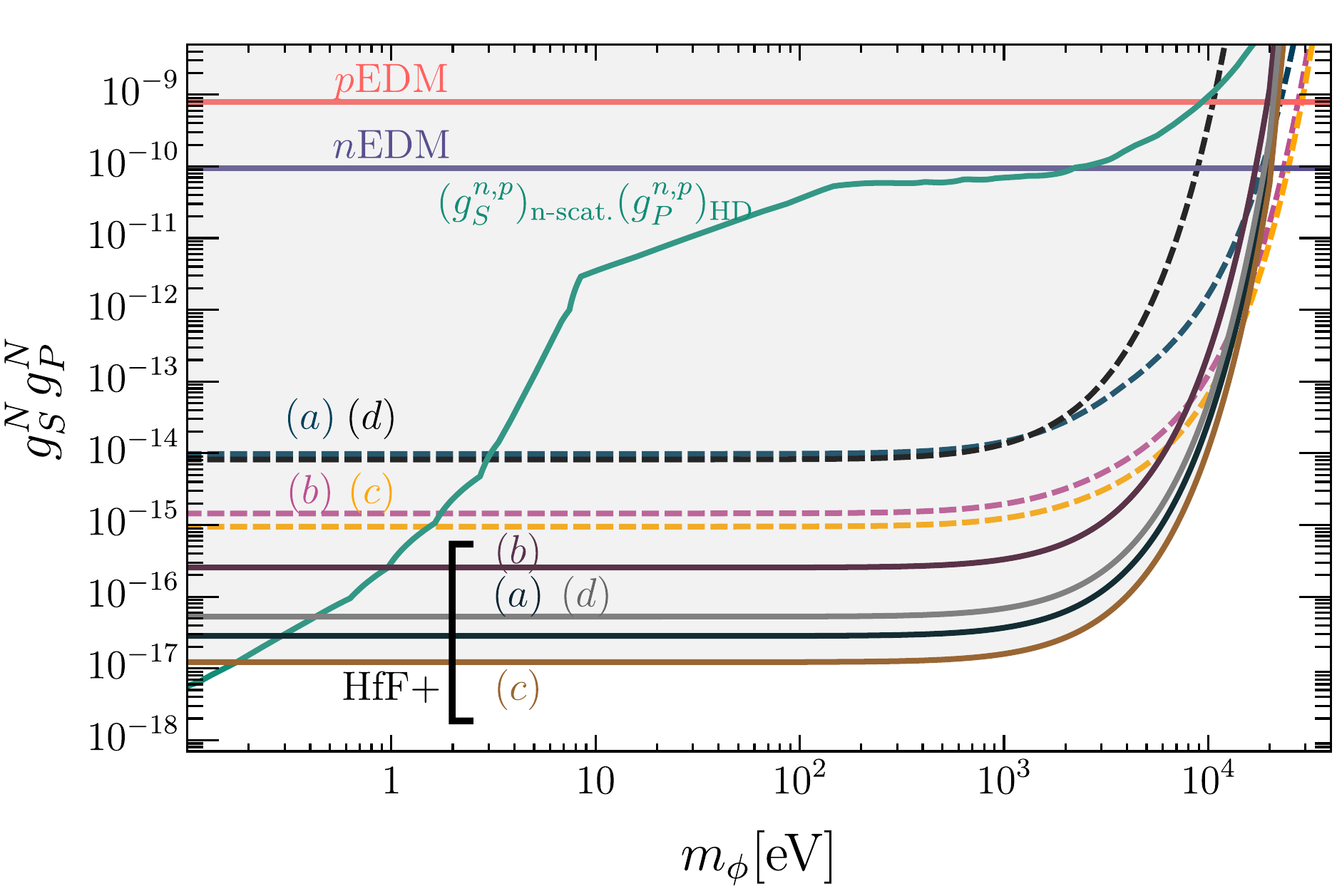}
    \caption{The constraints on $g^{p/n}_S g^{p/n}_P$ that can be set with a measurement with aggressive precision of $0.001\Hz$ of rovibrational transitions in \molecule. We show the bounds for the four benchmark scenarios \caseA,\caseB,\caseC and \caseD. We show the bounds that can be set by reinterpreting the $e$EDM measurement in HfF$+$ as well as combined existing bound from neutron scattering and vibrational spectroscopy in HD.  }
    \label{fig:TV_bound}
\end{figure}

Here we do an order-of-magnitude estimation as to the sensitivity of precision hyperfine measurements in \molecule{} to probe TV new physics. 
We consider the purely nucleon-nucleon interaction and leave an analysis of the other T-violating sources to future work.
The SM contribution to TV or CPV is negligibly small.
External electric and magnetic fields are required to orient the molecule such that $\expval{\vec{\sigma}\cdot\hat{r}}\neq 0$. 
We can choose a transition such that all nuclear spins are oriented along the direction of the magnetic field. 
For simplicity, we take $\vec{\sigma}=(0,0,1)$ for all nuclei. 
The bound obtained, for the four cases \caseA, \caseB, \caseC{} and \caseD{} are 
\begin{align}
    g_S^p g_P^p &\leq 9.8 \times 10^{-13} \nonumber \\
    g_S^n g_P^n &\leq 1.4 \times 10^{-13} \nonumber \\ 
    g_S^{p,n} g_P^{p,n} &\leq 9.5 \times 10^{-14} \nonumber \\
    g_S^{p,-n} g_P^{p,-n} &\leq 8.2 \times 10^{-13}\,,
\end{align}
for $m_\phi \lesssim \eV[k]$ and assuming an accuracy of $0.1 \Hz$. 
The mass dependent bounds are shown in Fig.~\ref{fig:TV_bound}. 
For the bound to be competitive with bounds set using diatomic \eEDM{} bounds we need a precision of the measurement comparable to the \eEDM{} measurements, \ie, $10^{-5} \Hz$. 
This is because the TV effect in both systems has a similar absolute value, the nuclear charges are of the same order of magnitude and the polarization is comparable. 
We leave a detailed analysis of the optimal transition to choose until experiment reaches this limit. 
We comment that a measurement with such an accuracy may also be able to probe the \eEDM{} with a similar sensitivity as the current probes in diatomic molecules~\cite{Roussy:2022cmp,ACME:2018yjb}, but detailed calculations are required.

\section{Conclusion}
\label{sec:conclusion}

In this work, we have shown that chiral molecules provide a unique environment for probing PV new physics. 
For example, chiral molecules can be the most sensitive terrestrial probe of PV in the case of a new spin-1 particle that couples only to neutrons, or to neutrons and protons oppositely. 
The chirality switch of chiral molecules provides a straightforward approach to remove the majority of significant SM contributions, thus the proposed method is free of SM background and does not require sophisticated SM calculations.
For the case of coupling only to the proton, for $\Hz[m]$ precision of the PV energy difference measurement, we would improve current bounds.
When coupling to protons and neutrons evenly, our bound is less constraining than the combined bound due to neutron scattering and HD precision measurements.
We also show that measurements of the SM PV energy difference can constrain the electron-nucleon PV interaction, to the level of APV. 
The bound we extract from current data is weaker than the APV bounds.  
Additionally, we have shown that chiral molecules can probe TV new physics. 
They would present a competitive probe only when the precision of the measurement becomes comparable to that of \eEDM{} measurements in diatomic molecules.

As development of state preparation techniques for \molecule{} progresses, opportunities for precision metrology of properties other than the vibrational degrees of freedom, which are well motivated, will arise. 
This paper motivates the development of hyperfine and rotational precision spectroscopy in \molecule{} and chiral molecules in general.

\begin{acknowledgments}
We thank Arie Landau, Inbar Savoray and Gilad Perez for collaboration at early stages of this project.
We would like to thank Joshua Baraban, Itay Bloch, Yevgeny Stadnik and Stephan Schiller for useful discussions. 
We thank Dima Budker and Inbar Savoray for comments on the manuscript. 
C.B. and Y.So. are supported by grants from the NSF-BSF (grant No. 2021800), the ISF (grant No. 483/20), the BSF (grant No. 2020300) and by the Azrieli foundation.
P.B.C. is supported by the U.S. NSF (grant No. PHY-2110489).
Y.Sh. gratefully acknowledges funding by the European Union (ERC, 101116032 – Q-ChiMP).
Views and opinions expressed are however those of the authors only and do not necessarily reflect those of the European Union or the European Research Council Executive Agency. Neither the European Union nor the granting authority can be
held responsible for them. YSh. also acknowledges support from the US-Israel Bi-national Science Foundation (BSF) [Grant no. 2022160] and Israel Science Foundation (ISF) [Grant no. 1142/21].
\end{acknowledgments}

\appendix 
\section{Quantum chemical calculations} 
\label{app:Qcomp}

The structural and electronic properties of CH$_2$BrI$^+$ and CHDBrI$^+$ were computed with coupled cluster theory including single, double, and perturbative triple excitations [CCSD(T)]~\cite{Raghavachari1989:CCSDt,Bartlett1990:CCSDt} as implemented in the CFOUR quantum chemistry code~\cite{Matthews2020:cfour,CFOUR}. 
The optimized geometry and first-order electronic properties were calculated using analytic gradient techniques~\cite{Watts1992:uhfCCSDt_grad}, while spin-rotation parameters were calculated within the effective one-electron spin-orbit operator approach of Ref.~\cite{Tarczay2010:spinrot} based on analytic second-derivatives for open-shell CCSD(T)~\cite{Szalay1998:uhfCCSDt_hessian}. 
Because this implementation of spin-rotation constants in CFOUR is only available with non-relativistic CCSD(T) and without effective core potentials, the all-electron 6-311G* and 6-311G** basis sets~\cite{Krishnan1980:6311Gxx,Curtiss1995:6311Gxx,Glukhovtsev1995:G2} were chosen for all geometry and property calculations. 

Preliminary calculations of fine and hyperfine properties were performed for the planar CH$_2$Br and CH$_2$I radicals to assess the performance of the non-relativistic CCSD(T) predictions for molecules containing the heavy atoms Br and I. 
The calculated and experimental values for the spin-rotation ($\epsilon_{aa}$, $\epsilon_{bb}$, and $\epsilon_{cc}$), nuclear magnetic hyperfine ($a_F$, $T_{aa}$ and $T_{bb} - T_{cc}$), and nuclear electric quadrupole coupling constants ($\chi_{aa}$ and $\chi_{bb} - \chi_{cc}$) are summarized in Table~\ref{tab:ch2br_ch2i}. 
The fractional agreement for each of the large in-plane components of the spin-rotation tensor ($\epsilon_{aa}$ and $\epsilon_{bb}$) is reasonably good (1--10\%). 
The out-of-plane component ($\epsilon_{cc}$) has much worse fractional accuracy (30--40\%), but this behavior is typical of planar $\pi$ radicals like CH$_2$Br and CH$_2$I for which the out-of-plane spin-rotation parameter has a small absolute magnitude by symmetry arguments~\cite{Ozeki2007:ch2br,Bailleux2010:ch2i}. 
The magnetic and electric quadrupole hyperfine coupling constants are also in generally good agreement, with the largest fractional error for the iodine Fermi contact parameter ($a_F$). 
This discrepancy is again not surprising at this level of theory, which includes neither core-electron correlation nor orbital-following zero-point vibrational effects, both of which can be important for calculating accurate electronic spin densities in planar $\pi$ radicals~\cite{Changala2024:propargyl}. 
Nonetheless, this theoretical approach is shown to provide qualitative to semi-quantitative accuracy for the fine and hyperfine properties of Br and I-bearing radicals.

\begin{table*}
    \begin{tabular}{lrrrr}
    \hline 
     & \multicolumn{2}{c}{CH$_2$$^{79}$Br} & \multicolumn{2}{c}{CH$_2$I} \\ 
     Parameter & Calc.\footnotemark[1] & Expt.\footnotemark[2] & Calc.\footnotemark[1] & Expt.\footnotemark[2] \\ 
     \hline 
     $\epsilon_{aa}$        & $-$11151.5 & $-$12569.80(2) & $-$29031.1 & $-$29409.76(1) \\
     $\epsilon_{bb}$        & $-$633.8   & $-$699.23(7)   &   $-$910.5 & $-$926.284(2)  \\
     $\epsilon_{cc}$        & 90.6       & 68.39(7)       &      295.9 &    208.565(2)  \\
     \\[-3mm]
     $a_F$(Br/I)            &  26.3      & 22.7904(9)     & 5.4        & $-$15.8381(5)\\
     $T_{aa}$(Br/I)         &  $-$151.1  & $-$149.718(2)  & $-$165.1   & $-$152.9644(7) \\
     $T_{bb}-T_{cc}$(Br/I)  &  $-$287.2  & $-$365.68(2)   & $-$280.2   & $-$430.966(2) \\
     $\chi_{aa}$(Br/I)      &  516.8     & 518.339(4)     & $-$1631.7  & $-$1745.022(3)\\ 
     $\chi_{bb}-\chi_{cc}$(Br/I) &  $-$107.7  & $-$132.78(6)   & 319.1      & 434.20(1) \\ 
     \\[-3mm]
     $a_F$(H)               & $-$77.5    & $-$60.208(1)   & $-$75.1    & $-$57.6046(7) \\
     $T_{aa}$(H)            & $-$20.5    &  $-$21.894(3)  & $-$19.3    & $-$20.708(2)\\
     $T_{bb}-T_{cc}$(H)     & 29.6       & 28(3)          & 28.6       & 19.3(1)\\
     
     \hline 
    \end{tabular}
    \caption{The fine and hyperfine parameters (in MHz) of CH$_2$Br and CH$_2$I in their $^2$B$_1$ electronic ground states.\label{tab:ch2br_ch2i}}
    \footnotetext[1]{Analytic equilibrium values at the frozen-core UHF-CCSD(T) level of theory with the 6-311G* basis set.}
    \footnotetext[2]{Experimental values from Refs.~\cite{Ozeki2007:ch2br,Bailleux2010:ch2i}}
\end{table*}
%

The equilibrium geometry of the $^2$A$'$ ground electronic state of CH$_2$BrI$^+$ optimized at the UHF-CCSD(T)/6-311G** level of theory is listed in Table~\ref{tab:ch2bri_geo} in its principal axis system (PAS). 
Table~\ref{tab:ch2bri_rot} lists the rotational constants and permanent electric dipole moment components for CH$_2$BrI$^+$ and CHDBrI$^+$, Table~\ref{tab:ch2bri_aniso} lists the isotropic hyperfine coupling constants (which are the same for both isotopologues), and Table~\ref{tab:ch2bri_aniso} lists the spin-rotation tensor and anisotropic magnetic and electric quadrupole hyperfine coupling constants. 
For the magnetic and electric quadrupole hyperfine constants, it is straightforward to transform the values calculated directly for CH$_2$BrI$^+$ to those for CHDBrI$^+$ by rotating the tensors into the new rotational principal axis system. 
The transformation of the spin-rotation tensor requires a more elaborate procedure because it is a second-order property that depends on the rotational kinetic energy operator itself. 
Assuming that the spin-rotation values are dominated by second-order spin-orbit contributions, we use well known isotopic scaling relations~\cite{Brown1980:spinrot_iso} to derive the spin-rotation parameters of the deuterated isotopologue.


\bibliographystyle{utphys28mod}
\bibliography{chiral.bib}

\end{document}